\documentclass[preprintnumbers,preprint,showpacs,prd]{revtex4}
\input epsf
\newcommand{\be}{\begin{equation}}
\newcommand{\ee}{\end{equation}}
\newcommand{\ba}{\begin{eqnarray}}
\newcommand{\ea}{\end{eqnarray}}

\newcommand{\mC}{\mathcal C}

\def\NPB{{\em Nucl. Phys.} B }
\def\NPA{{\em Nucl. Phys.} A }
\def\PLB{{\em Phys. Lett.}  B}
\def\PRL{{\em Phys. Rev. Lett.} }
\def\PRD{{\em Phys. Rev.} D}

\begin{document}
\preprint{BU-HEPP-03-03}

\title{Lattice Results on the Connected Neutron Charge Radius}

\author{Alfred Tang\footnote{Present address: Max Planck Institut fur Kernphysik, Heidelberg,
Germany, D-69029} and Walter Wilcox}

\affiliation{Department of Physics, Baylor University, Waco, TX 76798-7316}

\author{Randy Lewis}

\affiliation{Department of Physics, University of Regina, Regina, SK S4S 0A2, Canada}

\begin{abstract}
We describe a calculation using quenched lattice QCD of the connected part of
the neutron electric form factor for momentum transfers in the range $0.3\, {\rm GeV^2} 
\stackrel{<}{\sim} -q^2 \stackrel{<}{\sim} 1.0 \,{\rm GeV^2}$. We extract the implied 
charge radius using a Galster parameterization and consider various ways of extrapolating the 
neutron charge radius value to the chiral limit. We find that the measured charge radii
may be reconciled to experiment by standard phenomenology and lowest or next to lowest
order contributions from chiral perturbation theory.
\end{abstract}


\maketitle

\section{Introduction}
There has been a surge of interest and activity in evaluating electromagnetic
form factors of nucleons, both on the experimental~\cite{exp} 
and theoretical (lattice) side~\cite{the}. Unfortunately, knowledge of 
the neutron electric form factor, $G^{n}_{e}(q^{2})$, has lagged behind. The extraction 
of this fundamental quantity is difficult because
of the dominance of the magnetic contribution in unpolarized measurements. 
In addition, free neutron targets do not exist making its measurement from 
deuteron targets prone to model dependent systematic errors of the order of 
$\sim 30\%$~\cite{bermuth}. However, new types of experiments with a polarized
electron beam and deuteron or $^3$He targets now lead to measurements of
$G^{n}_{e}(q^{2})$ with greatly improved accuracy~\cite{gao,bermuth}. 
  
There are reasons to believe that 
this form factor is the most interesting and revealing of the four nucleon 
electromagnetic form factors. The contribution of the various 
flavored sea and cloud quark components will no longer be hidden behind a huge valence
contribution, as is the case for the other electromagnetic form factors, and the sizes 
of the non-valence components could hold some surprises. Lattice QCD was invented to 
reliably sort out such subtle effects, and it will be a major accomplishment to 
predict/explain these properties.

The calculation described here is of the \lq\lq connected" part of $G^{n}_{e}(q^{2})$.
The electromagnetic current is a sum of the charge-weighted flavor
currents, each of which is a color singlet. Thus there are also quark 
self-contraction graphs, usually called \lq\lq disconnected" graphs, 
involving the various flavor currents which can also 
contribute to this quantity. (These self-contraction loops are of course connected
to the valence or cloud quarks by gluons.) We do not attempt to calculate these graphs, 
but will comment on the significance of our results for the disconnected
contribution in Section V.

Previous lattice calculations of nucleon electromagnetic form factors
using Wilson~\cite{one,two,three} and improved fermions~\cite{the} have had good
signals for the proton electric and magnetic form factors, $G^{p}_{e}(q^{2})$ 
and $G^{p}_{m}(q^{2})$, as well as the neutron magnetic form factor,
$G^{n}_{m}(q^{2})$, but noisy results for the neutron electric form 
factor, $G^{n}_{e}(q^{2})$. Similar to experiment, the lattice extraction of this
form factor is the most difficult. If lattice studies
are to keep pace with the improved experiments referenced above, we must move beyond the
qualitative stage in the calculation of this physical quantity.
A new, more precise, lattice calculation of $G^{n}_{e}(q^{2})$ is described here. 
The lattice used is larger ($20^{3}\times 32$),
more configurations ($100$) are employed than in Ref.~\cite{one}, and more 
sophisticated analysis methods are used.

Although we will 
calculate the electric form factor at several nonzero momentum transfers, 
we prefer at this point to concentrate on the implied charge radius, a zero 
momentum quantity which should allow a more reliable chiral extrapolation.
We will see that the neutron charge radius is a useful laboratory 
in which the physics of the quenched approximation can be studied, 
and its shortcomings quantified. One of the effects of quenching
on our results comes through the double pole \lq\lq hairpin" graphs. 
We will consider these graphs separately and provisionally conclude 
that these do not have a significant effect on our extrapolations. 
We think it extremely likely that the quenched approximation
gives a good representation of the connected part of $G^{p}_{e}(q^{2})$.
The quenched, connected simulations quoted above have yielded phenomenologically 
acceptable chirally extrapolated electromagnetic form factors for both 
mesons and baryons. The quenched approximation has also been 
generally successful in evaluating other quantities such as, for example, 
scalar and axial nucleon form factors~\cite{others}. The validity of this 
approximation, however, can not be assured and subtle dynamical/disconnected 
effects may remain. We will suggest further investigations which should help 
form a more complete physical picture in Section V.

\section{Numerical Details}

The periodic gauge configurations used in this study were generated from the standard one-plaquette
action at $\beta=6.0$ on $20^3\times 30$ lattices. These lattices were slightly enlarged
to $20^3\times 32$ by copying each time edge to the opposite side because of
an inverter requirement that at least one of the dimensions be an integral multiple of 8~\cite{wliu}.
The Cabibbo-Marinari pseudo heatbath was used and configurations were separated by 
at least 1000 sweeps. No significant auto correlations were observed in the real or imaginary
parts of the nucleon correlators. 100 configurations were used in the charge radius analysis.

We used $\kappa=0.150 ,0.152, 0.153,$ and $0.154$ in the standard
Wilson quark action. The pion masses used are

\begin{center}
$m_{\pi}a=0.578(2),\,\, 0.4772^{+9}_{-2}$~\cite{iwasaki},\,\, $0.4237(8)$~\cite{gock}, 
\,\, $0.364(1)$~\cite{allton} \\
\end{center}
\noindent
respectively. The nucleon masses, previously reported in Ref.~\cite{randy} on 2000 
configurations, are
\begin{center}
$m_{N}a=0.997(7),\,\, 0.869(2),\,\, 0.799(2), \,\, 0.728(3)$ \\
\end{center}
\noindent
(The pion and nucleon masses at $\kappa=0.150$ were measured here on $t$=15 to 18
single exponential fits, where t=0 is the time origin of the quark propagators.)
The lattice spacing is taken to be the same as in Ref.~\cite{randy},
\begin{center}
a = 0.1011(7) fm
\end{center}
\noindent
obtained in Ref.~\cite{gock} from the physical string tension. The authors of 
Ref.~\cite{three} used the physical nucleon mass instead to arrive at $a=0.115(6)$fm. 

The electric form factors were measured from point source neutron two and three point
functions with the ratio given in Ref.~\cite{one}, with zero and non-zero
momentum two point functions evaluated at timestep $t'=9$:

\begin{equation}
G_{e}(q^{2})=\left(\frac{2E}{E+m}\right)^{1-\frac{t_{1}}{t'}}
\left(\frac{G_{nJ_{4}n}(t_{2},t_1;0,-\vec{q},\Gamma_{4})}
{G_{nn}(t_{2};0,\Gamma_{4})}\right)
\left(\frac{G_{nn}(t';0,\Gamma_{4})}
{G_{nn}(t';\vec{q},\Gamma_{4})}\right)^{\frac{t_{1}}{t'}}.\label{eq:ge}
\end{equation}

The time ordered two point function, using the standard neutron
interpolation field, $\chi_{\alpha}^{n}(x)= \epsilon^{abc}\psi_\alpha^{(d) a}(x)\psi_\beta^{(d) b}(x)
(\tilde{C})_{\beta\gamma}\psi_\gamma^{(u) c}(x)$, where $\tilde{C}=C\gamma_5$ and $C$ is 
the charge conjugation matrix, is (understood $\alpha$, $\alpha'$ sums)
\begin{equation}
G_{nn}(t;\vec{p},\Gamma)  \equiv 
 \displaystyle{\sum_{\vec{x}}} 
	e^{-i\vec{p}\cdot\vec{x}}	
\Gamma_{\alpha'\alpha} 
\langle{\rm vac}|T\left(\chi_{\alpha}^{n}(x)  
	\overline{\chi}_{\alpha'}^{n}(0)\right)|{\rm vac}\rangle ,
\label{eq1}
\end{equation}
where $\Gamma_{4}\equiv \frac{1}{2}\left(\begin{array}{cc}
       I & ~~~~0 \\
       0 & ~~~~0
\end{array}\right) $ in the $4\times 4$ Dirac space. The three point function we need, 
which uses the lattice conserved vector current, $J_{\mu}(x)$, is ($\vec{q} = \vec{p} - \vec{p}'$)

\begin{eqnarray}
	 &G_{nJ_{\mu}n}(t_{2},t_{1};\vec{p},{\vec{p}\,'},\Gamma)  
	  \equiv 
	-i\displaystyle{ \sum_{\vec{x}_{2},\vec{x}_{1}}} 
	e^{-i \vec{p}      \cdot\vec{x}_{2}}	
        e^{i\vec{q}\cdot\vec{x}_{1}} 
	 \Gamma_{\alpha'\alpha} \langle{\rm vac}|
		T\left(\chi_{\alpha}^{n}(x_{2})
	J_{\mu}(x_{1})\overline{\chi}_{\alpha'}^{n}(0)\right)|{\rm vac}\rangle.\label{3p}
\end{eqnarray}
The time position of the conserved charge density operator, $J_{4}(x_{1})$, which is extended one
lattice link in the time direction, is associated with the midpoint of that link, which means
half-integer values for $t_1$.

\section{results}

In Figs.1(a) and (b) we show an example of the raw data fits, in this case for $\kappa=0.152$ and
the lowest momentum transfer, where the time axis represents $t_1$ in Eq.(\ref{eq:ge}).
In order to establish reliable source positions for maintaining a good signal, we originally did the
calculations on 50 configurations with point nucleon sources at time steps 6 and 27. This is the $\Delta
t =21$ data referred to in Fig.1(b). The (b) part of the figure shows that a good plateau is already
forming at time step 7, although error bars are large. This allowed us to move the final source position
to time location 21, giving $\Delta t =15$ between sources, and three point correlation data at time
steps 7 - 10 relative to the origin were fit. The lowest three momentum transfers had similar plateaus
and form factor values were extracted for the other
$\kappa$ values. Fig.2 illustrates the form factors found at $\kappa=0.154$ and $0.153$; numerical
values and error bars are given in Table I. The error bars on the form factors only appear relatively
large because of the small central values. Comparing results at $\kappa=0.154$, the error bars here are
a factor of approximately 5 times smaller than the similar calculation in Ref.~\cite{one}. (They are also
significantly smaller than error bars given in Ref.~\cite{three}, although individual ${\kappa}$ results
are not given.) With the adopted scale for $a$, our smallest four-momentum transfer range on these
measurements is about $0.3 {\rm GeV^2}$, about half that used in Ref.~\cite{one}.

Before we go on to the charge radius fits, let us explain some of the philosophy of our fits of
the electric form factor data. A well known phenomenological form for $G^{n}_{e}(q^{2})$ is the Galster
parameterization~\cite{galster},
\begin{equation}
G^{gal}_{e}(q^{2}) = -\frac{\tau \mu}{1+p\tau}G^D(q^2),
\end{equation}
where $\tau\equiv -q^2/4m_N^2$, where $m_N$ is the nucleon mass and
\begin{equation}
G^D(q^2) = \frac{1}{(1-\frac{q^2}{m_D^2})^2},
\end{equation}
is the dipole form factor using the dipole mass, $m_D$. The most often used value of the parameter $p$
is 5.6 and $\mu\sim -1.91$ is the neutron magnetic moment. The neutron squared charge radius, given by
\begin{equation}
r^2_n = 6 \frac{dG^{n}_{e}(q^{2})}{dq^2}|_{q^2=0},
\end{equation}
implied by the Galster form is $r^2_n = 3\mu/2m_N^2= -3.24$ GeV$^{-2}$, which is only about $9\%$
different from the experimental value, $r^2_n=- 2.982(6)$ GeV$^{-2}$~\cite{rpp}. This form is used to fit
the form factor data given in Table I. At each value of $\kappa$ we have three values of $G_e^n(q^2)$.
There are also three parameters in the Galster form: $p$, $\mu$, and $m_D$. It is essential to have at
least one degree of freedom in the fits so that error bars and $\chi^2$ goodness of fits can be defined.
Dipole masses for Wilson fermion fits of the proton electric form factor were given in Table VII of
Ref.~\cite{one} at $\kappa=0.154$ and $0.152$. These values are adopted as input and listed in GeV in
Table II, along with the nucleon masses given above. In addition, the values at $\kappa=0.153$ and
$0.150$ were interpolated from the $\kappa=0.154,0.152,0.150$ dipole masses given in Ref.~\cite{one}.
These were plotted as a function of dimensionless quark mass, defined as $\ln(4\kappa_c/3\kappa-3)$,
where we used the central value in $\kappa_c=0.157096(28)^{+33}_{-9}$~\cite{iwasaki}. A linear plus
constant fit produced an excellent interpolation with $\chi^2 = 0.025$. The final Galster parameters
for the four fits are listed in Table II and are shown in Fig.2. Again, the
error bars on the extracted -$r^2_n$ appear large, but even at the largest $\kappa$ the error bar is
about 6 times smaller than the experimental value for this quantity. The error bars could have been made
significantly smaller by fixing the $p$ value in the Galster fits, but we choose not to do this.

In our chiral fits we are extrapolating the neutron charge radius using the formulas in the Appendix
from heavy baryon chiral perturbation theory and parameters in Table III across our four $m_{\pi}^2$
values to the physical charged pion mass at $139$ MeV. In doing these extrapolations, it is necessary to
adopt values of $f_{\pi}$ and $\Delta$ (the octet-decuplet mass difference). Our point of view is that
we are extrapolating down toward the chiral limit from our lowest pion mass at $\kappa=0.154$, so our
values of $f_{\pi}$ and $\Delta$ are measured at this value~\cite{down}. With our value of $a^{-1}$, the
result of Table 4 of Ref.~\cite{allton} gives $\Delta=158$ MeV, and Table VIII of Ref.~\cite{me2} gives
$f_{\pi}=120$ MeV at $\kappa=0.154$ when the $a^{-1}$ value adopted here is supplied. (It is also
consistent with the interpolated $f^{PT}_{\pi}$ value at $\kappa=0.154$ from Table XX of
Ref.~\cite{iwasaki}.) We also adopt the tree level QCD values $D=0.8$~\cite{flores}, the axial vector
octet-octet coupling, and $|{\cal C}|=1.4$~\cite{butler}, the axial vector octet-decuplet coupling. These
last two quantities are fixed from phenomenology at the chiral limit, so we can not claim complete
consistency~\cite{point}. Of course, other extrapolation schemes can be used, and the first reference
in \cite{the} for example assumes some parameter values at the chiral limit and extrapolates upward in
$m_{\pi}^2$ toward their measured (isovector) charge radii and magnetic moments using other parameters.

Fig.3 shows two extrapolations of the -$r^2_n$ values as a function of $m_{\pi}^2$ using
these parameters. One curve in Fig.3 assumes only quenched octet intermediate states, labelled \lq\lq O"
in Table IV. ($C_1$, $C_2$ give the analytic contribution to -$r^2_n$ and are defined in the
Appendix.) Perhaps surprisingly this produces an excellent fit of the data. Although the charge radii
values are initially small, they extrapolate to a physical charge radius which is consistent within
errors with the measured experimental result because of the chiral logs. The other curve in Fig.3, the
\lq\lq O+D" result from Table IV, gives the result for adding the leading octet and decuplet
contributions. As one can see this lowers the extrapolated -$r_n^2$, which is still consistent with
experiment. The error bars on the charge radius values at the chiral limit come almost completely from
the uncertainty in the $C_1$ values from Table IV, since the physical pion mass is essentially at the
chiral limit. The uncertainties in $C_1$ and
$C_2$ (given by $\pm 0.43$ GeV$^{-2}$ and $\pm 0.35$ GeV$^{-4}$, respectively) are independent of the fit
(as is the uncertainty in the extrapolated -$r_n^2$, as listed in the Table) since in the Marquardt
algorithm of Ref.~\cite{marq} the uncertainties are determined by either analytical or numerical
derivatives of the functional form with respect to the adjustable parameters. Since these are just
constant plus linear fits in $m_{\pi}^2$, the fixed nonanalytic part does not affect the uncertainties in
$C_1$, $C_2$ or the extrapolated -$r_n^2$.

Fig.4 contrasts the \lq\lq O+D" fit found in Fig.3 with two changed forms which also involve both the
octet and decuplet. One of these is the \lq\lq cutoff" form suggested in Ref.~\cite{detmold} in the
context of parton distribution functions, labelled as \lq\lq (O+D)(c)" in Table IV. The other form is the
higher order (in $1/m_{heavy}$) nonanalytic calculation of Ref.~\cite{puglia}, adapted to this quenched
situation and labelled \lq\lq (O+D)(h)" in Table IV. The figure illustrates that these forms can lower
the extrapolated -$r_n^2$ and can have a significant impact ($\sim 10-30\%$) upon the chiral limit, which
we think is a reasonable estimate of the systematic extrapolation error involved. The \lq\lq (O+D)(h)"
extrapolation raises the value from the \lq\lq (O+D)" fit and essentially hits the experimental value
exactly with a lowered $\chi^2_r$. It would be wrong to make too much of this agreement, but clearly the
fits we are making are consistent with the lattice data and the experimental value for the neutron
squared charge radius.

\section{Hairpin Considerations}

We have considered the quenched QCD contributions to the neutron and proton charge
radius from the so-called hairpin graphs using the methods of \cite{labrenz}. The
leading nonanalytic parts of these diagrams are proportional to $\delta \ln({m_{\pi}^2/\mu^2})$ where
$\delta \equiv m_0^2/(48\pi^2f_{\pi}^2)$ (for $m_u=m_d$) and $m_0$ is the usual hairpin \lq\lq mass"
parameter. These diagrams are the direct analog of the magnetic hairpin graphs considered recently in
Ref.~\cite{savage}. (See Figs. 2(a) and (b) of this reference for the relevant diagrams. The
$\alpha_{\phi}$ interaction should not contribute to a leading log for magnetic moments or charge radii
because of the two extra derivatives in the coupling.) Remarkably, we find that the contributions from
these graphs vanish identically for the proton and neutron. There are three sets of such graphs plus
wavefunction renormalization. The first set consists of a hairpin correction of the tree-level octet
charge radius vertex. When this is combined with the wavefunction renormalization, these two
contributions cancel identically, not only for the proton and neutron, but also for the other octet
baryons~\cite{comment}. In addition, there is a contribution at this order to the charge radius from the
electric quadrupole operator~\cite{bss}. It connects octet and decuplet baryon lines similarly to the
transition magnetic moment operator in Ref.~\cite{savage}. It vanishes for the proton and neutron
(although not for the other octet baryons) for the same reason that the magnetic transition operator
does; namely, that the flavor-charge coefficient vanishes. In addition, there are one-loop hairpin
contributions from the tree-level decuplet vertex, which also includes a wavefunction renormalization
part. These do not cancel, but vanish separately for the proton and neutron, again because of the
flavor-charge factor~\cite{decuplet}.

Unfortunately, there are also two-loop graphs which contribute at the same order as the one loop
graphs within chiral perturbation theory. This can be understood simply from the physical dimensions
of the charge radius, which goes like (mass)$^{-2}$. To lowest order in the hairpin interaction the
modification of the tree-level charge radius is of order $\delta (r^2_n)_{tree}$. However, a term of
the same order with the usual chiral perturbation theory factor $1/(4\pi f_{\pi})^2$ replacing the
factor of $(r^2_n)_{tree}$ may also be generated from a two-loop graph involving a primitive
electromagnetic interaction. Two-loop graphs have been considered recently in the context of mass
corrections in Ref.~\cite{mcgovern}. These graphs are clearly related to the graphs we need since
one of the pion loops may be changed to a hairpin loop just by differentiation with respect to the
pion mass and changing the flavor factor. Although we have stopped short of their evaluation
we can conclude from power counting that these charge radius graphs are logarithmically divergent
in the loop momentum and can in principle contribute to the chiral log charge radius coefficient.

An estimate shows that the contribution of any such hairpin chiral logs
in the proton and neutron charge radii should be small. Recently, there have been estimates of the value
of $\delta$, the hairpin coefficient, from lattice simulations~\cite{deltas}. For concreteness we will
take $\delta = 0.2$. As an example of the expected magnitude of the effect of the delta interaction on
the chiral log coefficient, let us take the magnitude of the ratio of the wavefunction renormalization
contribution (canceled by the modified tree-level piece) to the quenched chiral coefficient for the
proton or neutron:

\begin{equation}
R \equiv (\delta(r^2_n)_{tree} (D-F)^2)/(\frac{1}{(4\pi f_{\pi})^2}\frac{10}{3}D^2).
\end{equation}
Our fits below (see the $C_1$ coefficients in Table IV) give $(r^2_n)_{tree}< 1$ GeV$^2$, so that
for $D= 0.8$, $F= 0.5$~\cite{flores}, we obtain at most $R\sim 0.05$. If we replace the $(r^2_n)_{tree}$
by  another factor of $1/(4\pi f_{\pi})^2$, which would happen at two loops, we get the same estimate. Of
course, the flavor-charge coefficient here is specific to this interaction. However, barring an
exceptionally large coefficient it is likely that the effect of hairpins on
the chiral extrapolations will be negligible.

\section{Significance of Results}

We have seen that one can come remarkably close to understanding the value of the neutron
charge radius using quenched lattice QCD. Our calculations use the Wilson action and therefore
are limited to rather large pion masses. Although standard values for the $D$ and $\mC$ coefficients were
used in the analysis, there is nothing to force the quenched values of these coefficients to be the same
as in the full theory. The values we used should be considered merely typical, based on
standard phenomenology. However, our results for -$r^2_n$ are consistent with the expected rise from
the quenched chiral logs using these coefficients, and the fits to the data are excellent.
Of course lattice calculations, both quenched and dynamical, probing further into the chiral limit would
be helpful to verify this scenario.

We find that the higher order nonanalytic contributions in Ref.~\cite{puglia} or
the possible \lq\lq cutoff" form suggested in Ref.~\cite{detmold} can make a small but
significant impact on the chiral extrapolations. According to the argument in Section IV, we do not
expect hairpin contributions to make a significant contribution to the chiral log terms. However, we can
not rule out the possibility of large coefficients entering and this remains a slim possibility. An
exact evaluation of this quantity would be very welcome.

Of course, the disconnected part, which consists of up, down
and strange quark loops, should be added to our results before we compare to experiment.
Ref.~\cite{three} gives results for combined u, d, and s quarks and finds an increase of perhaps 0.5-1.0
GeV$^{-2}\approx .019 -039$ fm$^2$ in $-r_n^2$. (We are using the smallest $-q^2$ result from Fig.5(b)
of this paper and the Galster parameterization.) The question naturally arises as to whether our
results are compatible with such values. The fact that one finds good agreement with experiment using
standard phenomenology and only the connected part of the amplitude argues that the disconnected
contribution must be relatively small, again assuming the hairpin contribution is minor. An increase by
the amount suggested in Ref.~\cite{three} could perhaps be accommodated by our data, but likely at the
expense of the excellent fits found here. The present results illustrate the importance of
establishing a reliable set of extrapolation parameters, which can only be done by extending lattice
calculations for a variety of quantities closer to the chiral limit. For now, we conclude from our
lattice study that the connected part of the quenched amplitude is capable of explaining the bulk of
the neutron electric charge radius.

Note added: An independent discussion of charge radii in
quenched and partially quenched ChPT has just appeared~\cite{arndt}.

\section{Acknowledgements}

This paper has had a prolonged gestation period and represents the
final outcome of the preliminary results reported in Ref.~\cite{me}. A number
of computer systems at the National Center for Supercomputing Applications
were employed, including the CM2, CM5, Exemplar, Power Challenge, and Origin 2000 
computers. The production runs on the CM5 utilized the fast conjugate gradient written by
Weiqiang Liu~\cite{wliu}. The project was suggested to WW by Franz Gross and has been contributed 
to by Stephen Naehr, Phillip Kalmanson, and Nooman Karim under the Baylor University 
Research Experiences for Undergraduates Program.  AT was supported by the Baylor
University Postdoctoral Program. WW thanks Michael Ramsey-Musolf and  K. F. Liu for
helpful conversations and communications. It has been supported under NSF grants PHY-9203306, 9401068,
9722073, and 0070836 and in part by the Natural Sciences and Engineering Research Council of Canada. The
Baylor University Sabbatical program is also gratefully acknowledged.

\newpage
\begin{center} 
{\bf Appendix A: Quenched Chiral Charge Radii Expressions}
\end{center}

We use the $f_\pi \approx 93 {\rm MeV}$ normalization. For completeness we give
the lowest order chiral expressions for both octet baryon charge radii as well as
magnetic moments. The charge-flavor factors, $b^{(X)}_N$, are listed in Table III.

\be
(\delta\mu)_B=\frac{m_N}{(4\pi f_\pi)^2}\left(-\frac{\pi}{2}\sum_{X=\pi,K}b^{(X)}_Bm_X
\right)
\ee

\be
(\delta r^2)_B=\frac{1}{(4\pi f_\pi)^2}\sum_{X=\pi,K}b^{(X)}_B\left(-{5\over8}
\ln\left(\frac{m_X^2}{\mu^2}\right)-\frac{7}{8}\right)
\ee

\be
(\delta\mu)_T=\frac{m_N}{(4\pi f_\pi)^2}\sum_{X=\pi,K}
b^{(X)}_T\left(\frac{2}{3}\,F_m(m_X,\Delta,\mu)+\frac{10\Delta}{9}\right)
\ee

\be
F_m(m_X,\Delta,\mu)\equiv \left\{ \begin{array}{ll}
-\Delta \ln\left(\frac{m_X^2}{\mu^2}\right) + 2\sqrt{m_X^2 
-\Delta^2}\cos^{-1}\left( \frac{\Delta}{m_X} \right) & \mbox{,\,\,$m_X >\Delta$} \\
-\Delta \ln\left(\frac{m_X^2}{\mu^2}\right) - 2\sqrt{
\Delta^2 - m_X^2}\ln(\frac{\Delta+\sqrt{\Delta^2-m_X^2}}{m_X}) & \mbox{,\,\,$\Delta > m_X$}
\end{array}  \right.\label{fm}
\ee

\be
(\delta r^2)_T=\frac{1}{(4\pi f_\pi)^2}\sum_{X=\pi,K}
b^{(X)}_T \left(-\frac{5}{3}F_r(m_X,\Delta,\mu)-\frac{26}{9}\right)
\ee

\be
F_r(m_X,\Delta,\mu)\equiv \left\{ \begin{array}{ll}
\ln\left(\frac{m_X^2}{\mu^2}\right) + \frac{2\Delta}{\sqrt{m_X^2 
-\Delta^2}}\cos^{-1}\left( \frac{\Delta}{m_X} \right) & \mbox{,\,\,$m_X >\Delta$} \\
\ln\left(\frac{m_X^2}{\mu^2}\right) + \frac{2\Delta}{\sqrt{
\Delta^2 - m_X^2}}\ln(\frac{\Delta+\sqrt{\Delta^2-m_X^2}}{m_X}) & \mbox{,\,\,$\Delta > m_X$}
\end{array}  \right.\label{fr}
\ee
Note the sign typo in Eq.(6) of Ref.~\cite{puglia} in the $\Delta > m_X$ branch of
$F_r(m_X,\Delta,\mu)$. Although these quenched expressions do not appear in the literature, it is clear
that they are implied by the results in Ref.~\cite{kubis} for the full QCD charge radii and the
results of Ref.~\cite{savage} for the quenched magnetic moments since one can simply take the ratio of
the quenched to full QCD coefficients in Table III of Ref.~\cite{savage}, and apply them to the full QCD
charge radii results in \cite{kubis}. (One must also drop the $D$, $F$ independent terms from Table 1
of \cite{kubis}.) We have also independently verified the above results for charge radii and magnetic
moments. We have left analytic parts in the expressions so that others may independently verify these
results using dimensional regularization.

The functional form of the nonanalytic expressions used in the chiral extrapolations of the
neutron $-r^2_n$ are given below. \\
\noindent
Octet only:
\be
{\rm O} =- \frac{1}{(4\pi f_\pi)^2}   
{10\over3}D^2\ln\left(\frac{m_{\pi}^2}{\mu^2}\right).
\ee
\noindent
Octet plus decuplet:
\be
{\rm O+D} = \frac{1}{(4\pi f_\pi)^2}\left\{   
-{10\over3}D^2\ln\left(\frac{m_{\pi}^2}{\mu^2}\right) +\frac{5}{6}{\cal C}^2F_r(m_{\pi},\Delta,\mu)
\right\}.
\label{o+d}
\ee
\noindent
Octet plus decuplet cutoff form:
\be
{\rm (O+D)(c)} ={\rm O+D}+\frac{1}{(4\pi f_\pi)^2}\left\{   
{10\over3}D^2\ln\left(\frac{\Lambda^2+m_{\pi}^2}{\mu^2}\right) -
\frac{5}{6}{\cal C}^2F_r(\sqrt{m_{\pi}^2+\Lambda^2},\Delta,\mu)
\right\}.
\label{cut}
\ee
\noindent
Octet plus decuplet with higher order ($\sim 1/{\rm heavy^3}$) correction:
\begin{eqnarray}
{\rm (O+D)(h)} = \frac{1}{(4\pi f_\pi)^2}\left\{   
-{10\over3}D^2\left(\ln\left(\frac{m_{\pi}^2}{\mu^2}\right) + 
\frac{23\pi m_{\pi}}{24M_N} \right) \right. \\   
\left. +\frac{5}{6}{\cal C}^2\left(F_r(m_{\pi},\Delta,\mu) -\frac{2}{M_N}
G(m_{\pi},\Delta,\mu)\right)\right\},
\nonumber
\end{eqnarray}
where (Ref.~\cite{puglia})
\be
G(m_{\pi},\Delta,\mu)\equiv \left\{ \begin{array}{ll}
\Delta\ln\left(\frac{m_{\pi}^2}{\mu^2}\right) + \frac{2\Delta^2 - m_{\pi}^2}{\sqrt{m_{\pi}^2 
-\Delta^2}}\cos^{-1}\left( \frac{\Delta}{m_{\pi}} \right) & \mbox{,\,\,$m_{\pi} >\Delta$} \\
\Delta\ln\left(\frac{m_X^2}{\mu^2}\right) + \frac{2\Delta^2-m_{\pi}^2}{\sqrt{
\Delta^2 - m_{\pi}^2}}\ln(\frac{\Delta+\sqrt{\Delta^2-m_{\pi}^2}}{m_X}) & \mbox{,\,\,$\Delta > m_{\pi}$}
\end{array}  \right.\label{g}
\ee
\noindent
To all such forms for $-r^2_n$ are added the analytic terms $C_1 + C_2 m_{\pi}^2$ where $C_1, C_2$ are
constants. Note that there can be additional terms depending on $\Lambda$ in the \lq\lq cutoff" form,
Eq.(\ref{cut}), but we include only the modifications to the nonanalytic terms in Eq.(\ref{o+d}) in the
same spirit as Ref.~\cite{detmold}.

\newpage

\begin{table}[h]
\caption{Numerical results on neutron form factors
at four values of $\kappa$ and various values of the four momentum 
transfer $-q^2$.}
\begin{center}
\begin{tabular}{c|c|c}
 $\kappa$ & $\,\,-q^{2}\,\, {\rm (GeV^{2})}$ & $\,\,G_e(-q^{2})$  \\
\hline
 & 0.358  & 0.0115(34)  \\
$0.154$ & 0.689 & 0.0134(43) \\
 & 0.996 & 0.0640(52)  \\
\hline
 & 0.361 & 0.0088(21)  \\
$0.153$ & 0.699 & 0.0099(29) \\
 & 1.02 & 0.0054(36)  \\
\hline
 & 0.364 & 0.0067(15)  \\
$0.152$  & 0.707 & 0.0076(21) \\
 & 1.03 & 0.0046(27)  \\
\hline
 & 0.366 & 0.0042(8)  \\
$0.150$  & 0.717 & 0.0049(13) \\
 & 1.05 & 0.0034(17)  \\
\end{tabular}
\end{center}

\caption{Galster fit parameters. The values of $p$ and $-r^2_n$ are
results of the fits, the values of $m_{N}$ and $m_{D}$ are input.
$\chi^2$ is the chi-squared of the fit.}
\begin{center}
\begin{tabular}{c|c|c|c|c|c}
 $\kappa$ &\,\, $m_{D}$ {\rm (GeV)}& \,\,$m_{N}$ {\rm (GeV)}& $\,\,p$ &
 $\,\,-r^2_n\,\,{\rm (GeV^{-2})}$ & $\chi^2$
\\
\hline
$0.154$ & 1.16 & 1.42 & 23.(36)  & 0.65(51) & 0.62\\
\hline
$0.153$ & 1.24 & 1.56 & 30.(42)  & 0.48(34) & 0.50\\
\hline
$0.152$  & 1.32 & 1.69 & 33.(42)  & 0.34(20) & 0.41\\
\hline
$0.150$  & 1.47 & 1.94 & 38.(39)  & 0.18(7) & 0.27\\
\end{tabular}
\end{center}
\end{table}

\begin{table}
\caption{Charge-flavor factors $b^{(X)}_N$ of baryon $N$ and Goldstone boson
$X$ of the quenched non-hairpin 1-loop diagrams.  The symbols $B$ and $T$
denote the octet and decuplet baryons respectively.}
\begin{center}
\begin{tabular}{c|cc|cc}
$N$ & $b^\pi_B$ & $b^K_B$
& $b^\pi_T$ & $b^K_T$  \\
\hline
$p$ & ${16\over3}\,D^2$ & 0 & $-{1\over2}\,\mC^2$ & 0 \\
$n$ & $-{16\over3}\,D^2$ & 0 & ${1\over2}\,\mC^2$ & 0 \\
$\Sigma^+$ & 0 & ${16\over3}\,D^2$ & 0 & $-{1\over2}\,\mC^2$ \\
$\Sigma^0$ & 0 & ${8\over3}\,D^2$ & 0 & $-{1\over4}\,\mC^2$ \\
$\Sigma^-$ & 0 & 0 & 0 & 0 \\
$\Lambda$  & 0 & $-{8\over3}\,D^2$ & 0 & ${1\over4}\,\mC^2$ \\
$\Xi^0$    & 0 & $-{16\over3}\,D^2$ & 0 & ${1\over2}\,\mC^2$ \\
$\Xi^-$    & 0 & 0 & 0 & 0
\end{tabular}
\end{center}
\end{table}

\begin{table}
\caption{Final fit parameters on some chiral extrapolations of $-r^2_n$.
The parameters in the fits are: $\Delta=158$ MeV, $f_{\pi}=120$ MeV, $D=0.8$, $|{\cal C}|=1.4$, $\mu=1$
GeV, $\Lambda =550$ MeV. \lq\lq O" denotes octet only; \lq\lq O+D" denotes octet plus decuplet; \lq\lq
(O+D)(c)"  means the cutoff form, \lq\lq (O+D)(h)" means the higher order octet plus decuplet form.
$\chi^2_r\equiv \chi^2/N_{DF}$ is the reduced chi-squared of the fit, where the number of degrees of
freedom, $N_{DF}=2$.}
\begin{center}
\begin{tabular}{c|c|c|c|c}
 fit &\,\, $C_{1} {\rm (GeV^{-2})}$ & \,\,$C_{2} {\rm (GeV^{-4})}$ &
 \,\,$-r^2_n\,\,{\rm (GeV^{-2})}$ & $\chi^2_r$
\\
\hline
O & -0.230 & 0.501 & 3.49(43)  & 0.0008\\
\hline
O+D & 0.060 & -0.093 & 2.49(43) & 0.010\\
\hline
(O+D)(c) & 0.580 & -0.373 & 2.22(43) & 0.022\\
\hline
(O+D)(h) & 0.321 & 0.252 & 2.90(43) & 0.003\\
\end{tabular}
\end{center}
\end{table}

\newpage

\begin{figure}
\begin{center}
\epsfxsize=5.5 in
\epsfbox{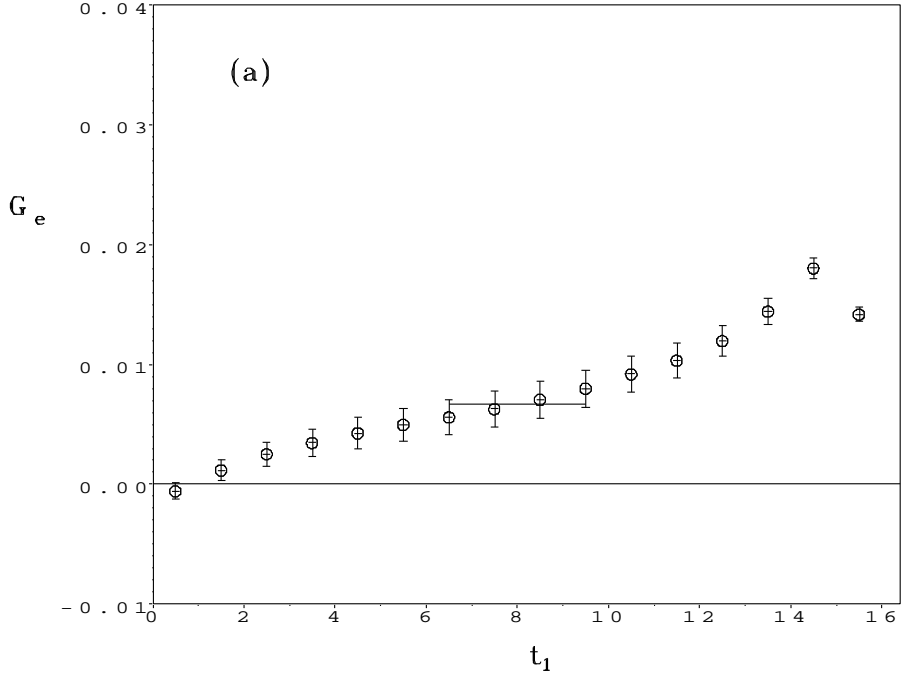}
\vskip .5cm
\epsfxsize=5.5 in
\epsfbox{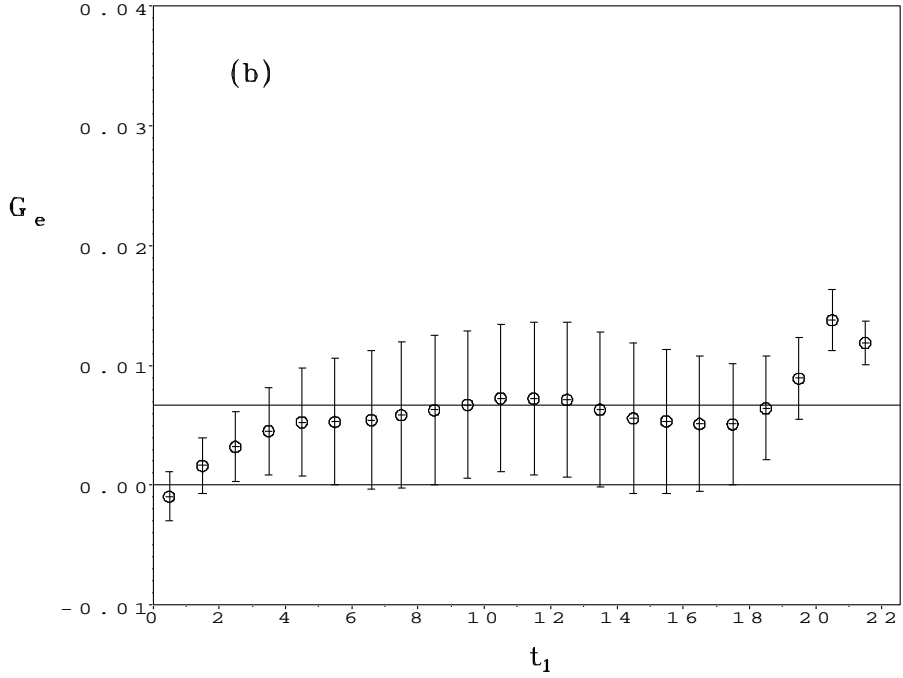}
\caption{(a) $\kappa=0.152$ electric form factor data for $p= \pi/10$ on the source separation
$\Delta t=15$ lattice for 100 configurations. (b) Similar graph on the source separation $\Delta t =21$
lattice for 50 configurations. The fit value on the $\Delta t =15$ data is shown superimposed on the
$\Delta t =21$ data in the (b) part.}
\vskip .5cm
\end{center}
\end{figure}

\newpage

\begin{figure}
\begin{center}
\epsfxsize=5.5 in
\epsfbox{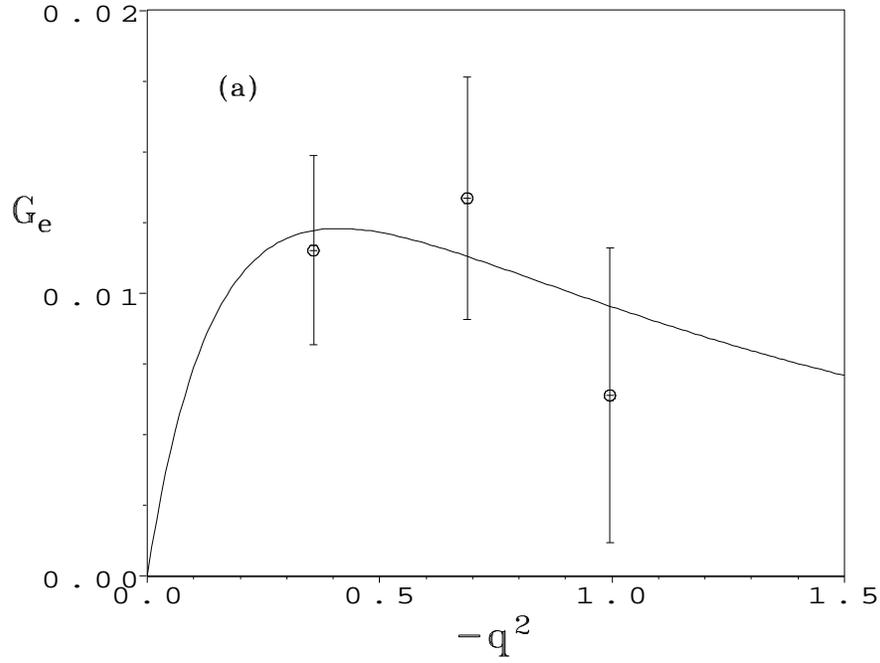}
\vskip .5cm
\epsfxsize=5.5 in
\epsfbox{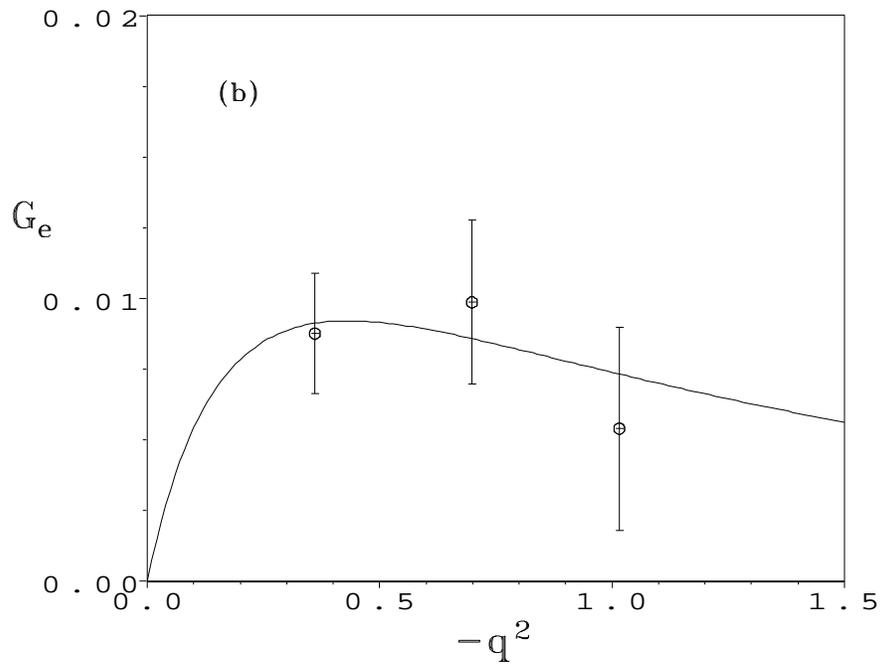}
\caption{(a) $\kappa=0.154$ neutron electric form factor data compared to the Galster fit
as a function of $-q^2$ in GeV$^2$. 
(b) Similar graph for $\kappa=0.153$.}
\vskip .5cm
\end{center}
\end{figure}

\newpage

\begin{figure}
\begin{center}
\epsfxsize=7.0 in
\epsfbox{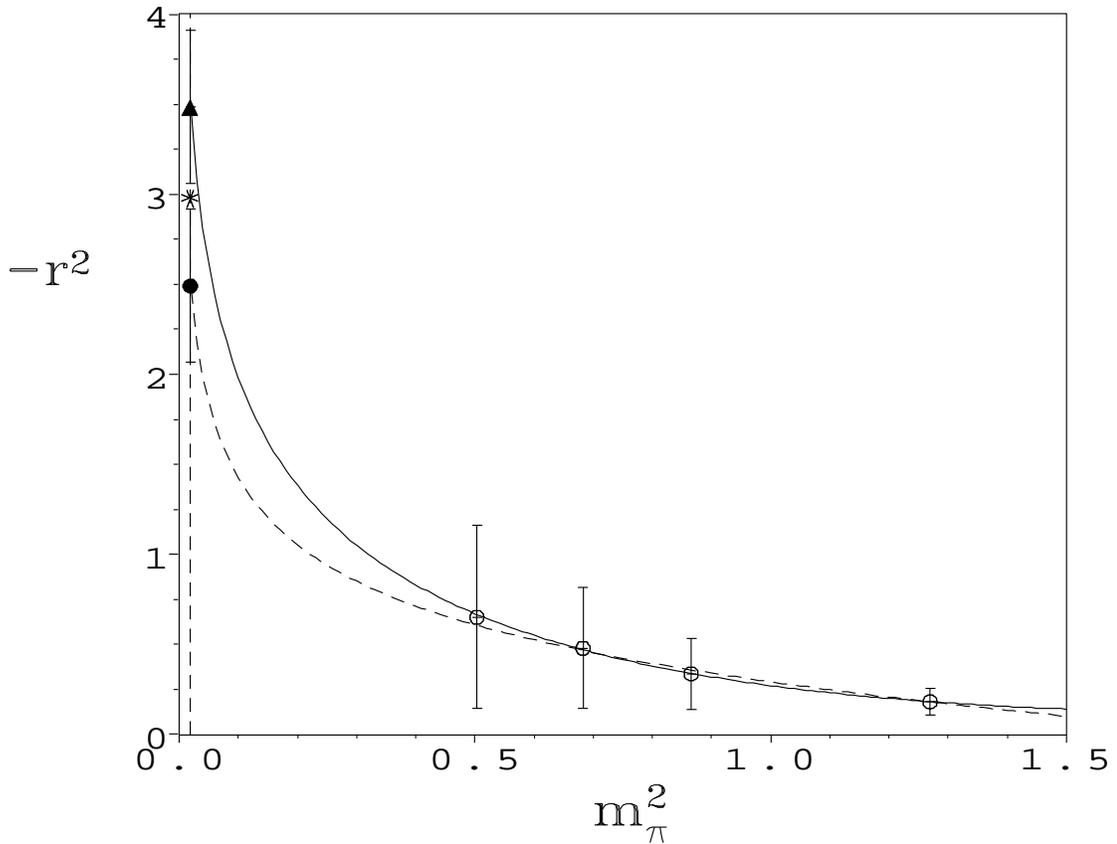}
\vskip .5cm
\caption{Two chiral extrapolations of the neutron charge radius data (in ${\rm GeV}^{-2}$)
as a function of pion mass squared (in ${\rm GeV^2}$).
The solid line is the pure quenched octet (\lq\lq O" in Table IV) extrapolation, the 
dashed line gives the octet plus decuplet (\lq\lq O+D"). The vertical dashed line represents
the physical pion mass squared and the burst on this line represents the
experimental charge radius. The lattice data are represented by open circles and
are given as a function of squared pion mass in ${\rm GeV^2}$. Error bars are
given on the extrapolated values (triangle - octet, solid circle - octet plus
decuplet) at the physical pion mass squared.}
\end{center}
\end{figure}

\newpage

\begin{figure}
\begin{center}
\epsfxsize=7.0 in
\epsfbox{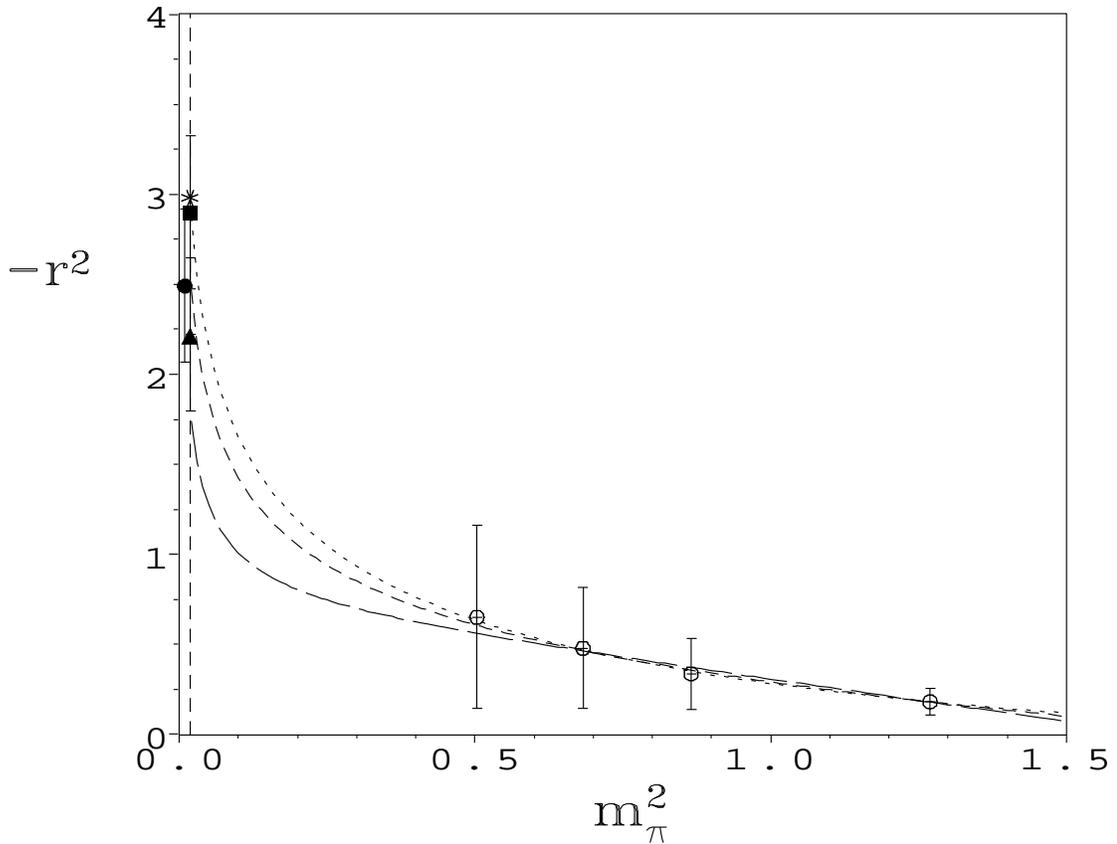}
\vskip .5cm
\caption{Three chiral extrapolations of the neutron charge radius data (in ${\rm GeV}^{-2}$) as a
function of pion mass squared (in ${\rm GeV^2}$). From top to bottom (shortest dashed line to longest).
All fits take into account both the octet and decuplet. The top line gives the result of the
higher order fit (\lq\lq (O+D)(h)" in Table IV), the middle is the (lowest order) octet plus decuplet
(\lq\lq O+D") also shown in Fig.3, and the lowest dashed line is the cutoff form (\lq\lq (O+D)(c)"). The
solid square, circle and triangle give the chiral values of these three fits, respectively. The solid
circle has been moved to the left for clarity of observations of the various error bars.
See Fig. 3 for meanings of the other symbols.}
\end{center}
\end{figure}


\begin{thebibliography}{99}
\bibitem{point} The results in Table IV of Ref.~\cite{randy}, which represent a combined fit of
disconnected scalar and electromagnetic contributions to the nucleon form factors, resulted in a
slightly smaller value of $\mC\sim 0.9 -1.0$ at $\kappa=0.152$.
\bibitem{decuplet} Although the one-loop decuplet diagrams are conventionally included
in the discussions of leading chiral logs, one can actually see from Eqs.(\ref{fm}),
(\ref{fr}), and(\ref{g}) that the $m_X \rightarrow 0$ limit produces forms proportional to
$\ln(\Delta^2/\mu^2)$. That is, when $m_X$ vanishes the heavy baryon denominator structure $v\cdot l +
\Delta$, where $v$ is the velocity and $l$ is loop momentum, still has a fixed infrared
scale $\Delta$ and does not produce a chiral log in $m_X$.
\end{thebibliography}
\end{document}